# SURFACE TEMPERATURE ANALYSIS OF A HOT TITANIUM TARGET IN ARGON MAGNETRON SPUTTERING SYSTEM

Hacı Ahmedov[a,1], Aleksandr A. Kozin[b], Viktor I. Shapovalov[b], Aygerim N. Djandalieva[b] Arif Demir[c], Beste Korutlu[a]

[a]*TÜBİTAK National Metrology Institute, Gebze Yerleşkesi P.K. 54 41470 Gebze/Kocaeli, Turkey*

[b]*St. Petersburg Electrotechnical University, 197376, 5 Prof. Popov str., St. Petersburg, Russia*

[c]*Gedik University, Cumhuriyet Mahallesi İlkbahar Sokak No: 1-3-5 Yakacık 34876 Kartal/Istanbul Turkey*

**Abstract:** Target temperature in magnetron sputtering systems is a significant parameter which not only affects the physical and chemical properties of the deposited film but also stimulates the reactive sputtering process. Therefore, it is of great interest to establish a relationship between the target temperature and other parameters of magnetron sputtering process such as the gas pressure and the discharge power. This paper is devoted to the experimental and numerical studies on titanium target temperature in Argon medium. Numerical simulations for the target temperature are performed by solving the Heat equation for the magnetron sputtering system enclosed in Argon filled chamber. The simulations are compared with the experimental temperature data. The difference between the simulated results and experimental data on the power dependence of target temperature is explained by an empirical model of heat transfer from the ions to the gas.

[1]Corresponding author: haji.ahmadov@tubitak.gov.tr

## 1. Introduction

Sputtering of metal targets in different gaseous media is of constant interest in coating technologies since it is used as a tool for the synthesis of inorganic films with different chemical compositions [1–3]. The progress in sputtering technology has led to the origination of hot-target magnetrons where the target surface temperature could be heated up to its melting point [4–7]. In hot target magnetron sputtering, the film deposition rate increases significantly compared to the traditional cold target ones since the bounding energy of the atoms in the lattice decreases effectively when their thermal energy increases. Another distinguishing feature of hot target magnetron is that the heat emission from its surface results in an additional heating on the substrate which then could modify the crystalline structure of the film [8-10]. This may be advantageous in the synthesis of thin film with special mechanical, optical, electrical and chemical properties. Furthermore, in reactive sputtering, where a reactive gas is introduced along with an inert gas, the target temperature plays a significant role [11-14]. It is a practical application among the sputtering techniques and is a combined physical, electrical, and chemical process as the activated reactive gas forms chemical bonds with the sputtered atoms to form a new compound [15-18].

It is clear from aforementioned features of hot target magnetron that the target temperature is a prominent parameter to be determined. Most often, non-contact methods are used for measuring the target surface temperature via radiation spectra analysis [15, 18, 19]. However, it is difficult to achieve high accuracy via radiation spectra as the measured radiation contains blackbody radiation from the surface of the target along with the glow discharge radiation. Hence, software modelling is preferred to be used to determine the target temperature for given the gas pressure and the discharge current. The temperature analysis of target surface via software allows also the determination on the optimal dimensions of the magnetron based on the material type. Note that there are extensive studies in the literature on the modelling of magnetron sputtering processes both with cold targets [20-25] and hot targets [5, 10, 26]. In the software modelling, numerical analysis of the heat equation with suitable boundary conditions is performed. There are mainly four heat transfer mechanisms on the target surface defining its temperature: the incoming heat carried by the ions accelerated in the electric field between cathode and anode (ionic heat flux density), the heat transfer from the target surface to the refrigerator which is cooled by running water (metallic heat flux density), outgoing heat flux density due to the blackbody radiation emitted from surface of the target (radiation heat flux density) and outgoing heat flux density due to the collisions of Argon with the target surface (gaseous heat flux density). The first three heat transfer mechanisms in magnetron sputtering system with a hot titanium target are simulated via COMSOL Multiphysics Thermal Process Package in [27]. In this paper, we extend the results of this work by introducing the gaseous heat transfer mechanism in addition to the metallic, ionic and radiation ones. Although the impact of gaseous heat transfer mechanism is less compared to the first three ones, it allows us to gain better understanding of the effects of gas pressure and discharge power on the target temperature.

## 2. Experiment and Model

This section is divided into two subsections. The first subsection is devoted to experimental studies on target surface temperature while the second one is allocated to the corresponding simulations.

### 2.1. Experimental Studies on Temperature

Experimental studies on the hot target magnetron sputtering process were performed in a vacuum chamber in cylindrical shape with a diameter of 600 mm and a height of 400 mm. The chamber, filled with Argon gas, is equipped by high resolution spectrometer which is used for the target temperature measurements and magnetic flux density meter which is used for the characterization of magnetic flux density distribution on the target surface. The schematic cross-section of magnetron sputtering system and the vacuum chamber is given in

Fig. 1. The magnetron sputtering system powered by DC current source is composed of three parts (see Region I in Fig. 1). The upper part is the titanium target which has a disc shape with a diameter of 130 mm and thickness of 1 mm. The bottom part is the chromium cooling unit in the shape of a hollow cylinder with running water cooling. The height of the refrigerator is 4 mm. A magnetron gap between the target and the cooling unit is supplied by a titanium ring with the height of 1 mm. More detailed descriptions of the experimental set-up, the temperature and the magnetic field measurements can be found in [27].

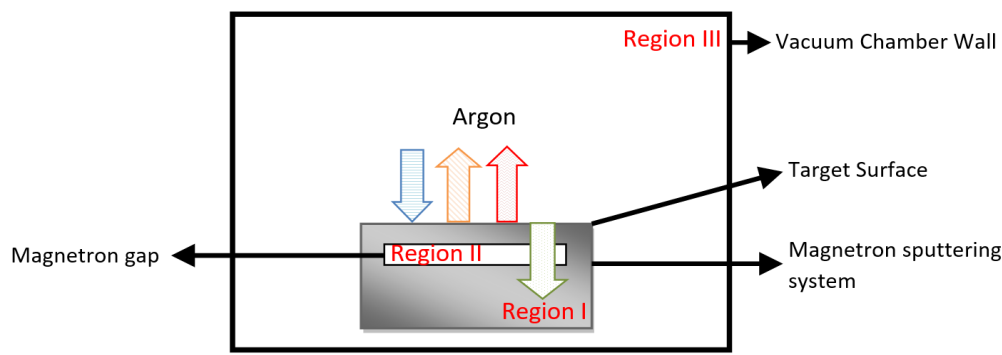

**Figure 1.** Schematic cross-section of 3D magnetron sputtering system inside the vacuum chamber. Given in stripped (blue) arrow is ionic heat flux, in twilled (orange) arrow is radiation heat flux, in wavy (red) arrow is gaseous heat flux and dotted (green) arrow is metallic heat flux.

Experimental data on discharge current versus temperature at 2 mTorr pressure is given in Fig. 2. The temperature is obtained by using the blackbody radiation spectra recorded by the spectrometer.

The current – voltage characteristics of the Argon discharge is obtained at 2 mTorr (see blue diamond data in Fig. 3) by measuring the voltage between the anode and the cathode. The discharge current versus discharge voltage data can be divided into two regions as increasing and decreasing discharge voltages. In the first region, the electric field is more dominant in generating electrons and ions between the plates whereas in the second region thermal electrons due to the high temperature of the target surface prevails. It is straightforward to obtain the discharge current versus discharge power plot by using the standard relation between power, voltage and current. The current-power characteristic of glow discharge at 2 mTorr is given as red circles in Fig. 3.

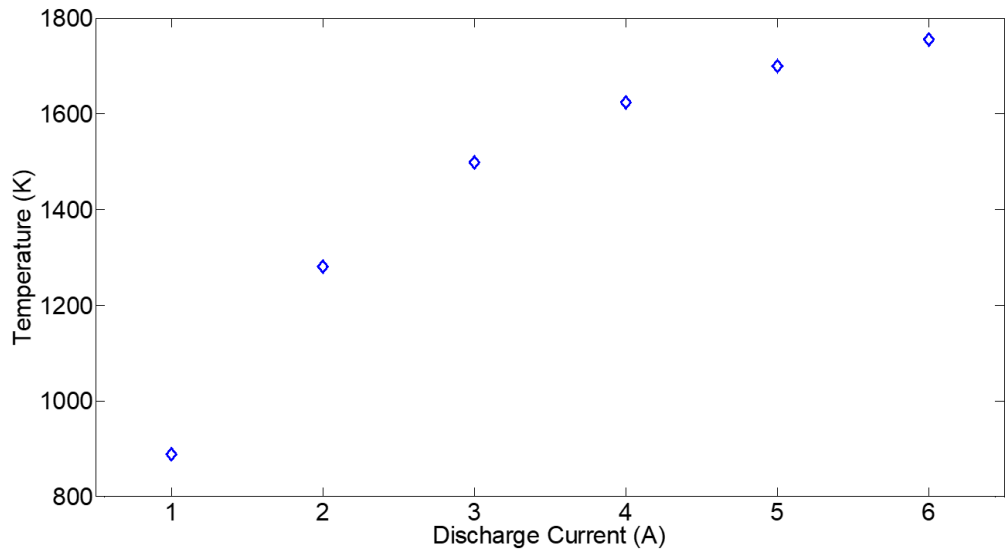

**Figure 2.** The target temperature with respect to the discharge current at 2 mTorr.

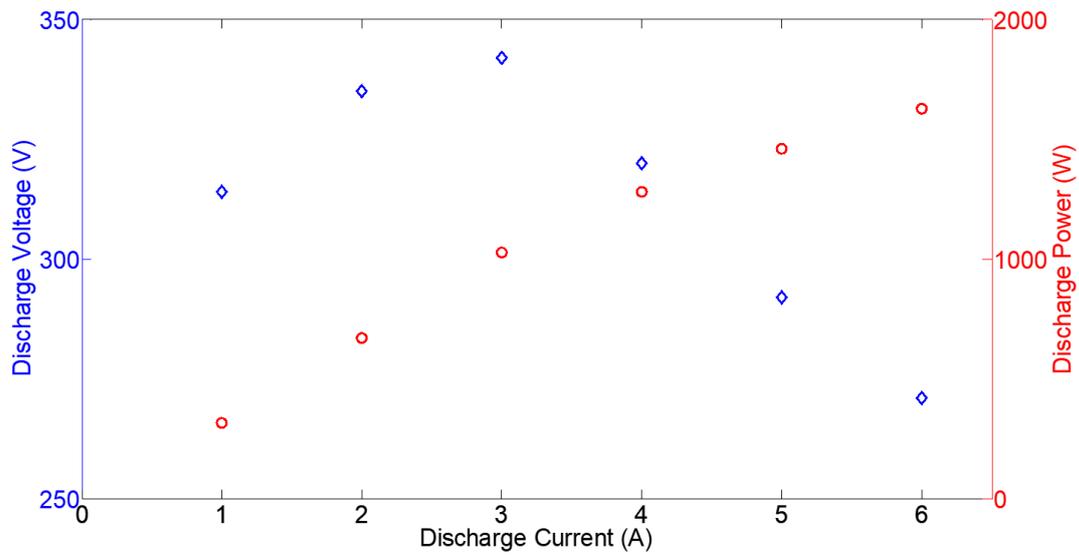

**Figure 3.** Given in blue diamond is the current - voltage characteristics of the glow discharge at 2 mTorr and given in red circles is the current - power characteristics of the glow discharge at 2 mTorr.

Target temperature with respect to the discharge power is given in Fig. 4 with the help of data in Fig. 2 and Fig. 3.

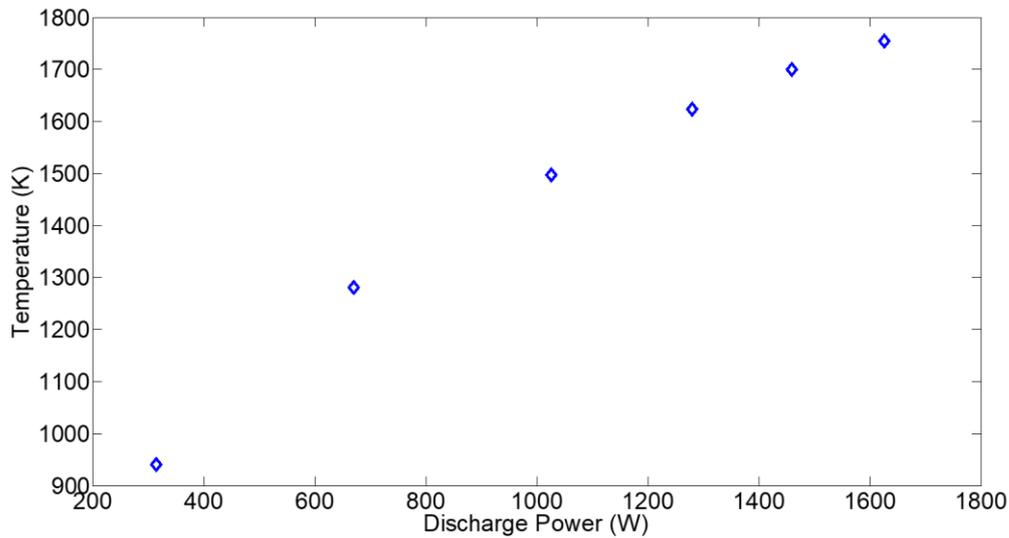

**Figure 4.** The target temperature with respect to the discharge current at 2 mTorr.

## 2.2. Temperature Simulation

In order to model the target temperature, one has to solve the stationary heat transfer equation in three separate regions inside the vacuum chamber: metallic region I which is the magnetron sputtering system, gaseous region II in the magnetron gap and gaseous region III between the magnetron sputtering system and the vacuum chamber. The gaseous and metallic heat flux densities are defined by the gradient of the temperatures in the relevant regions multiplied by the heat conductivity coefficient of the corresponding material with a minus sign. The radiation heat flux density in non-metallic regions are defined by the Pointing vectors of all harmonics emitted by metallic surfaces.

The Dirichlet boundary condition is applied on the vacuum chamber wall and at the bottom surface of the refrigerator where both the vacuum chamber and the refrigerator are treated as heat reservoirs with high heat capacities. In order to have consistent solutions in three different regions, one has to impose the heat flux balance condition on all remaining surfaces. The heat flux balance condition on the target surface is formulated such that the sum of the normal components of the metallic heat flux, the gaseous heat flux, the radiation heat flux and ionic heat flux densities is equal to zero. This is illustrated by arrows given in Fig. 1. In a similar manner, we write the heat balance equation for the other surfaces. The only difference is that the ionic heat flux density is equal to zero on those surfaces.

COMSOL Multiphysics Thermal Process Package is used in modelling the target surface temperature. The magnetic field of the magnetron is responsible from directing the ions towards the magnet. Therefore, the ionic heat flux density distribution on the target surface is modelled by using the magnetic flux density distribution data on the target surface. The two simulation parameters are the ionic heat flux and the gas pressure. The former is simply the integral of the ionic heat flux density over the target surface and the latter manifests itself through the dependence of thermal conductivity on the pressure. The simulation results on the target temperature with respect to the ionic heat flux at fixed pressures of 2 mTorr and 7 mTorr are given in Fig. 5.

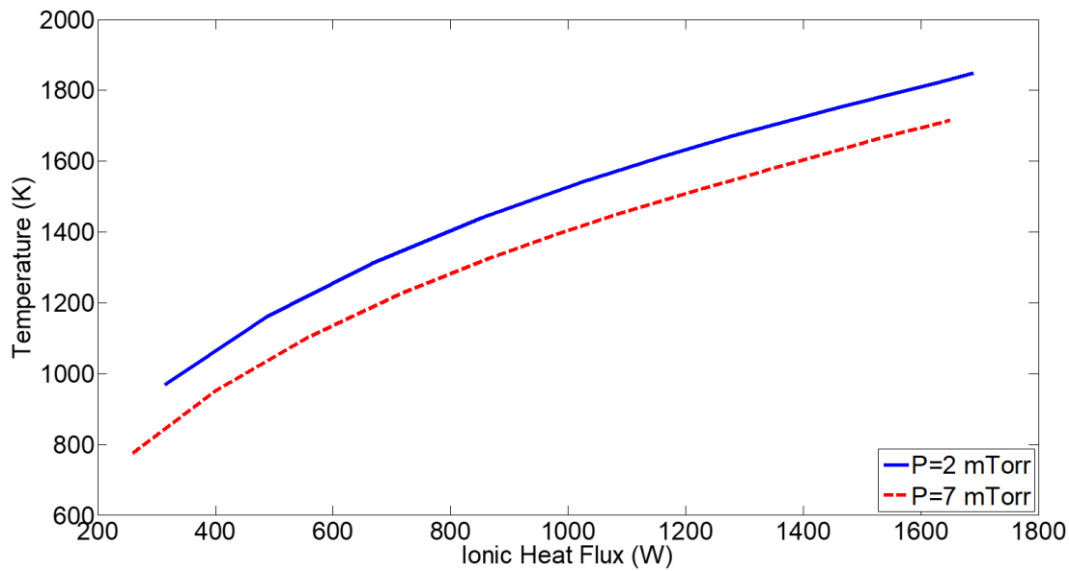

**Figure 5.** The simulated results on the target temperature with respect to discharge power at fixed pressures of 2 mTorr (blue solid-line) and 7 mTorr (red dashed-line).

### 3. Results and Discussion

In order to illustrate the validity of the model, the simulated results on the target temperature are compared with the corresponding experimetal ones. The experimental data (diamond) and simulation results (straight line) for power versus temperature at 2 mTorr are given in Figure

6. The differece between the simulation and experiment is due to the fact that not all of the discharge power is transferred to the ionic heat flux.

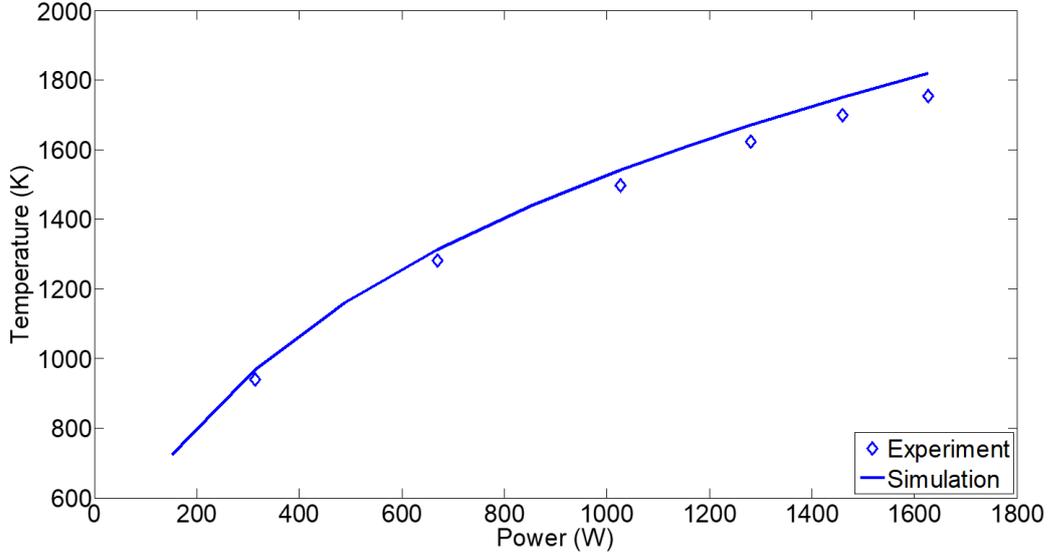

**Figure 6**. The experimental data and simulation results on target temperature with respect to power at 2 mTorr.

This difference between the discharge power and the ionic heat flux is called as the power loss. The power loss $\Delta p$ as a function of power is obtained by using first order Taylor expansion on the simulated target tempereture as a function of discharge power such that

$$\Delta p = \frac{\partial p}{\partial T_M}(T_M - T_E), \qquad (1)$$

where $T_M$ is the modelled target temperature and $T_E$ is the experimental target temperature. The power versus power loss plot is given in Fig. 7. It is clear from Fig. 7 that the more the power the more the power loss.

There are different sources for the power loss among which heat transfer from ions to gas is the leading one. Therefore, one should include the gas discharge in simulating the target temperature to shed light on the nature of this power loss. We are planning to study this in the forthcoming paper.

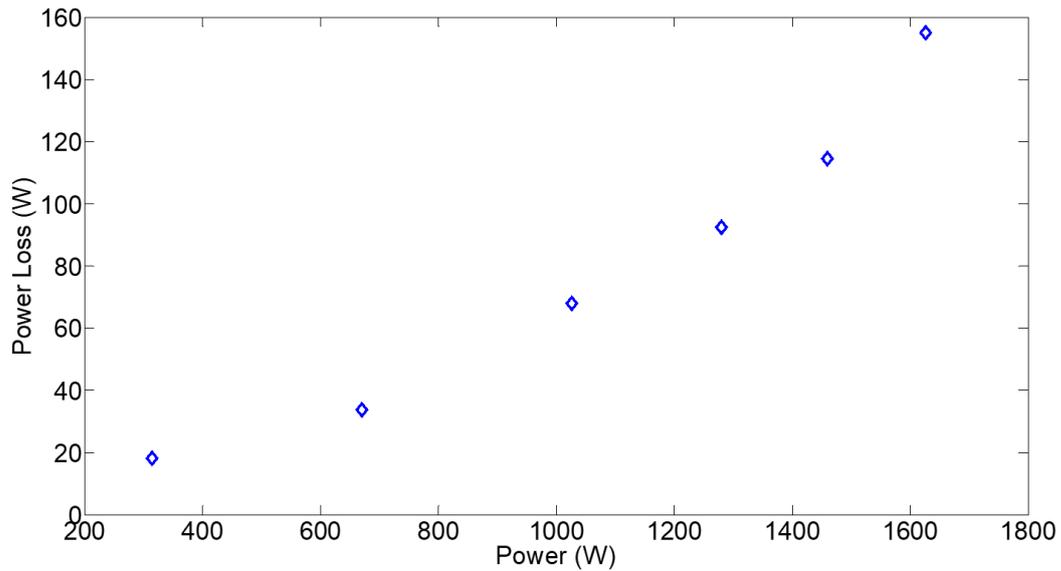

**Figure 7**. Power loss with respect to the power.

The simulation results on the target temperature with respect to the ionic heat flux are given in Fig. 8. The solid-line (blue) represents modelling without gas heat transfer and the dashed-line (red) represents modelling with gas heat transfer.

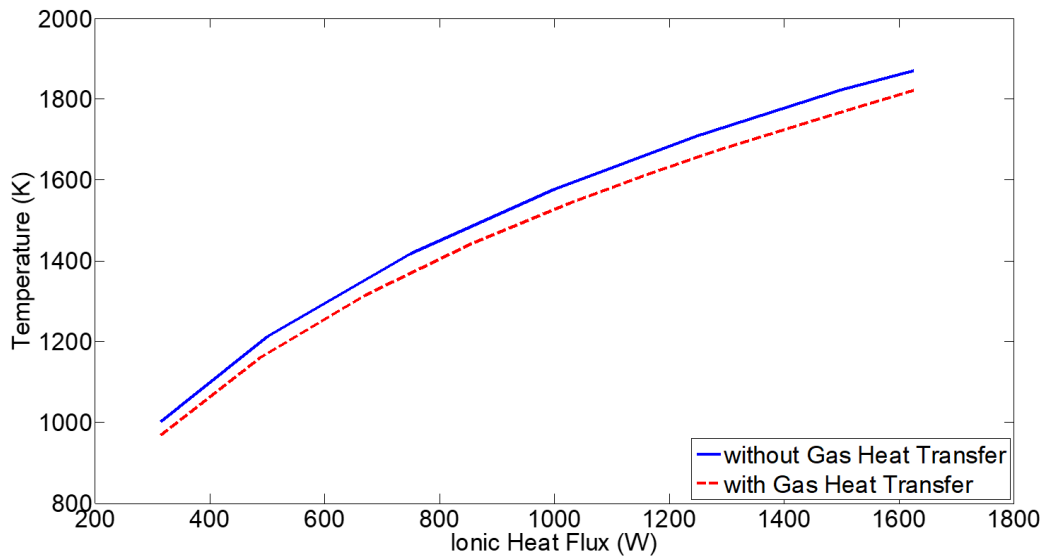

**Figure 8**. The simulation results on ionic heat flux with (red dashed-line) and without gas heat transfer (blue solid-line).

It is appearent from Fig. 8 that, gas heat transfer becomes more important process to be considered in modelling for increasing values of ionic heat flux since the thermal conductivity increases with temperature. It is important to note that, the metallic heat flux passes through the titanium ring while the the gaseous heat flux from all surfaces. Despite the thermal conductivity of titanium (20 W/m K) is much less compared to the other metals, it is still much more than the one of the gas. The impact of gas heat transfer mechanism increases with two effects: The first one is related to the corresponding surface areas. The metallic heat transfer from the target to the refrigerator occurs along the surface of target in contact with the titanium ring. The gaseous heat transfer, on the other hand, take place along both the upper surfaces of the target and the air gap. The second effect is gradient of the temperature from the surface of the target up to the refrigerator. The high temperature is concentrated at the center of the target and decreases strongly towards the edge of the target. Therefore the gradient of the temperaure in the ring is much smaller than the one in the gas, especially in the magnetron gap.

The linear dependence of the target temperature on the ionic heat flux prevails (see Fig.5). This is due to the fact that the main mechanism for heat removal from the target surface is metallic heat flux density transfering the heat to the refrigerator. Non-linearities in the ionic heat flux versus temperature curve are related to the radiation heat flux which is proportional to the fourth degree of temperature as it obeys to the Stefano-Boltzman law. The target temperature drops with increasing gas pressure (see Fig. 5). The reason for such a behaviour is that the thermal conductivity of the gas increases with an increase in its pressure.

**4. Conclusions**

The target temperature is an important parameter in the magnetron sputtering systems. Therefore, determining its dependence on the gas discharge parameters is of great interest. In the course of this work, the physical model for description of thermal processes for the magnetron sputtering system inside the vacuum chamber is proposed which includes the gaseous, the metallic, the radiation and ionic heat transfer mechanisms. Even though the impact of gaseous heat transfer in temperature distribution on the target surface might seem less at first glance, depending on the dimensions of the magnetron ring, it might contribute significantly. Essentially, the simulations are used both for practical reasons and for design purposes. Using the COMSOL Multiphysics thermal process package simulations of the titanium target temperature are performed for different values of the discharge power and the Argon pressure.

The simulation results and experimental data are compared to check the reliability of the model. The difference between the simulated results and experimental data on dependence of target temperature on the power can be clarified by modelling the heat transfer from the gas discharge to the gas. In this paper we constructed an empirical model for this difference.

Although, the result presented in the work is specific to Argon medium and the titanium target, the proposed model is universal. By necessary adjustments of parameters, it can be used to simulate target temperature for various magnetron sputtering configurations, target materials and mediums.

## Acknowledgement

The authors H. Ahmedov and B. Korutlu thank TÜBİTAK UME for their support in the study.